\documentclass[rnote]{aa}      

\usepackage{txfonts}
\usepackage{longtable}
\usepackage{natbib}
\usepackage{graphicx}

\bibpunct[, ]{(}{)}{;}{a}{}{,}
\usepackage{supertabular}

\usepackage{color}

\begin{document}

\title{Possible discovery of the r-process characteristics in the abundances of metal-rich barium stars}

\author{
  W. Y. Cui\inst{1,2},
  B. Zhang{1}, 
  J. R. Shi\inst{3}, G. Zhao\inst{2,3},
  W. J. Wang\inst{4}, \and
  P. Niu\inst{1}
}

\offprints{W.Y. Cui; \email{wycui@bao.ac.cn}}
\institute{Department of Physics, Hebei Normal University, 20 Nanerhuan East
Road, Shijiazhuang 050024, China \\
 \email{wycui@bao.ac.cn}
\and School of Space Science and Physics, Shandong University at Weihai, Weihai 264209, China\\
\and National Astronomical Observatories, Chinese Academy
of Sciences, 20A Datun Road, Beijing 100012, China\\
\and Baoding University, Baoding, Hebei 071000, China\\
}

\date{Received  / Accepted }

  \abstract{}
 {We study the abundance distributions of a sample of metal-rich barium stars  provided by Pereira et al. (2011)  to investigate  the s- and r-process nucleosynthesis in the metal-rich environment.
}
{We compared the theoretical results predicted by a parametric model with the observed abundances of the metal-rich barium stars.}
{We found that six barium stars have a significant r-process characteristic, and we divided the barium stars into two groups: the r-rich barium stars ($C_r>5.0$,  [La/Nd]\,$<0$)  and normal barium stars. The behavior of the r-rich barium stars seems more like that of the metal-poor r-rich and CEMP-r/s stars. We suggest that the most possible formation mechanism for these stars is the s-process pollution, although their abundance patterns can be fitted very well when the pre-enrichment hypothesis is included. The fact that we can not explain them well using the s-process nucleosynthesis alone may be due to our incomplete knowledge on the production of Nd, Eu, and other relevant elements by the s-process in metal-rich and super metal-rich environments  (see details in Pereira et al. 2011).}
{}


\titlerunning{Metal-rich barium stars}
\authorrunning{Cui et al.}

\maketitle
%
\section{Introduction}\label{Sect:intro}

Heavy elements ($Z > 30$) are produced by  neutron-capture-process nucleosynthesis, which is subdivided into rapid (r-) and slow (s-)  processes, depending on the competition between the $\beta$-decay and the next neutron-capture.  The main site of the s-process reactions is inside the low-mass stars \citep[$M<3.0$M$_\odot$,][]{gal98, str06} during their asymptotic-giant-branch (AGB) phase.  The r-process nucleosynthesis is always associated with exploding astrophysical events, such as core-collapse type II supernovae (SNII)  or neutron star mergers.  The accurate site (sites), however, is (are) still not confirmed for the r-process nucleosynthesis.

According  to \citet{bur00}, 85\% of Ba and 3\% of Eu in the solar system were contributed by the s-process. Thus, Ba and Eu are usually regarded as the representative elements of the s- and r-process, respectively. More than 80\% of the carbon-enhanced metal-poor (CEMP, [C/Fe]\,$>1.0$ and [Fe/H]\,$<-2.0$) stars exhibit excesses of the s-process elements (e.g. Ba),  that is, they are CEMP-s stars \citep{aok07}. Furthermore, about half of  the CEMP-s stars also show significant r-process characteristics (  e.g., Eu enrichment), CEMP-r/s stars  \citep{bee05}.  Different astrophysical sites  have been suggested in which the s- and r-process nucleosynthesis occurs, thus the existence of CEMP-r/s stars is confusing. Many mechanisms for their formation have been suggested,  but, none of them  was able to explain all of their observational characteristics  \citep[see details in ][]{jon06}. 

Barium stars  are a class of chemically peculiar stars  that were first  described  by \citet{bid51}. Barium stars 
show enhancements of carbon and  elements heavier than iron, which are often used to constrain the s-process nucleosynthesis theory \citep{all06, dra08}.  Based on  their evolution state, barium stars are divided into two  groups: barium giants and barium dwarfs. Barium giants are a group of chemically peculiar G and K giants  that have evolved into the red giant branch.   Fewer barium dwarfs than giants  have been confirmed \citep{gom97}. Barium giants have similar luminosities to red giants  \citep{sca76}, which are more luminous than barium dwarfs and less luminous than AGB stars in which the s-process nucleosynthesis is expected to occur \citep{gal98,bus01}. 
  Because of this,  the binary hypothesis was required to explain the abundance anomalies of barium stars. In this mechanism, a massive companion  that is now a white dwarf invisible in optical  observations first produced the s-materials during its AGB evolution phase, and then transferred  them to the observed star via  Roche-lobe overflow or wind accretion in a binary system.

\citet{per11b} reported detailed chemical abundances and kinematics of a sample of metal-rich barium stars. For convenience, we  divided the barium stars into two groups:   stars with a metallicity  lower than the solar value ([Fe/H]\,$<0$) 
  are referred to as metal-poor barium stars;  the other group of stars with [Fe/H]\,$>0$  is referred to as metal-rich barium stars. 
  Of the 12 stars of \citet{per11b}, 11 stars were confirmed as barium stars, including seven barium giants and four metal-rich CH subgiants.  In addition, \citet{per11a} pointed out that no metal-rich barium dwarfs were discovered. In the metal-rich sample, abundances of  the five neutron-capture  elements Y, Zr, La, Ce, and Nd were provided \citep{per11b}. 

In this paper, we analyze the abundance patterns of the neutron-capture elements of  the metal-rich barium stars reported by \citet{per11b} to investigate the s-process nucleosynthesis in the metal-rich environments. The paper is organized as  follows: Section 2 presents our calculation, results and discussion, and in Section 3 our conclusions are drawn.

\section{Results and  discussion}\label{Sect:results}

To investigate the s-process nucleosynthesis in metal-rich conditions, we  studied the barium star sample provided by \citet{per11b} using the parametric model for low-mass AGB stars \citep{zha06, cui07, cui10}. In our model, the contributions from r-process  are also considered. We obtained the theoretical abundance $N_{i}$ of the $i$th element  based on the following formula:
\begin{equation}
N_{i}(Z)=C_{s}N_{i,\ s}+C_rN_{i,\ r}10^{[Fe/H]},
\end{equation}
where $Z$ is the metallicity of the star, $N_{i,\ s}$ and $N_{i,\
r}$ are the abundances of the $i$th element produced by the s- and
r-process  (per Si $=10^6$ at $Z=Z_\odot$), $C_s$ and
$C_r$ are the component coefficients representing the contributions
of the s- and the  r-process. During the calculations, the observed abundances were used to constrain the physical parameters of the neutron-capture nucleosynthesis.

\begin{table*} [!ht]
 \caption{Calculation results of the physical parameters  for metal-rich barium stars.\label{tbl-1} }
 \centering
 \begin{tabular}{lccccccc}\hline\hline
Star &[Fe/H]& $\Delta\tau$ & $r$ & $\tau_0$ & $C_s$ & $C_r$ & $\chi^2$ \\
& &(mbarn$^{-1}$) & &(mbarn$^{-1}$) & & & \\\hline
  CD-25\degr 6606 &0.12&0.15&0.02  &0.04&0.00971&15.3&0.69281  \\
  HD 46040 &0.11&0.48&0.01  &0.10&0.00075&10.3&9.77175  \\
  HD 49841 &0.21&0.13&0.02  &0.03&0.03390&17.2&0.98713  \\
  HD 82765 &0.20&0.20&0.06  &0.07&0.00154&3.4&0.13097  \\
  HD 84734 &0.21&0.16&0.1  &0.07&0.00293&13.8&0.37039  \\
  HD 85205 &0.23&0.32&0.01  &0.07&0.00040&11.9&1.92897  \\
  HD 100012 &0.18&0.01&0.93  &0.14&0.00044&6.5&3.50993  \\
  HD 101079&0.10&0.25&0.01  &0.05&0.00082&7.5&0.23397  \\
  HD 130386&0.16&0.17&0.22  &0.11&0.00156&0.0&0.04935  \\
  HD 139660&0.26&0.23&0.01  &0.05&0.00167&8.4&0.22529  \\
  HD 198590&0.18&0.26&0.05  &0.09&0.00080&3.4&0.07676  \\
  HD 212209&0.30&0.12&0.01  &0.03&0.03774&4.0&0.07312\\\hline
\end{tabular}
\end{table*}

The abundance ratio of barium to europium is particularly sensitive to the relative contributions of the s- and r-process nucleosynthesis for the heavy elements formation. [La/Eu] and [Ba/Nd] are also used  as important indicators \citep{mas04,sim04}. Here, we used [La/Nd]  as indicator of the relative contributions of the s- and r-process nucleosynthesis  for the heavy element productions in a star. In  the solar system the heavy elements were formed  by the r- and s-process, but different elements have different proportions of the two neutron-capture processes. According to \citet{bur00}, the main s-process produced 75\%  of La and 47\%  of Nd in the solar system. La is usually regarded as one of the typical s-process  elements, while the Nd production from both the r- and s-process is similar in  the solar  system and the exact  value depends on  the AGB models \citep{bur00,arl99}.  Thus, different [La/Nd] ratios mean different r- and s-process contributions, and the lower  the [La/Nd] values, the  higher the r-process contributions.

Our results  are shown in Table~\ref{tbl-1}. The  low $\chi^2$ values  imply that our theoretical predictions for the yields of neutron-capture elements fit the observational abundances well for the corresponding metal-rich barium stars. This supports the binary hypothesis for the  formation of barium stars even in metal-rich environments.  To confirm this, however,  long-term radial-velocity monitoring for these metal-rich barium stars is encouraged. HD 46040 is an exception, which has a  high $\chi^2$ value, {which 9.77175, this means that the theoretical results  cannot fit the observed abundances well. The reason may  be its unusually high enrichment  of neutron-capture elements especially for La and Ce  compared with other stars with similar metallicities. \citet{per11b} suggested that HD 100012 does not belong to the group of barium stars because of its low [s/Fe] value, which is similar to the observed values of some normal field stars at this  metallicity; here, 's' represents the mean abundance of Y, Zr, La, Ce, and Nd.  However, the low [s/Fe] value of HD 100012 is mainly due to its low Y  abundance; in fact, the other four elements still show  enrichment compared with other normal stars  \citep[][see their Figure 8]{per11b}. Thus, this star is still included in our discussion. 

\begin{figure*}[!ht]
  \centering
  \includegraphics[width=1\textwidth]{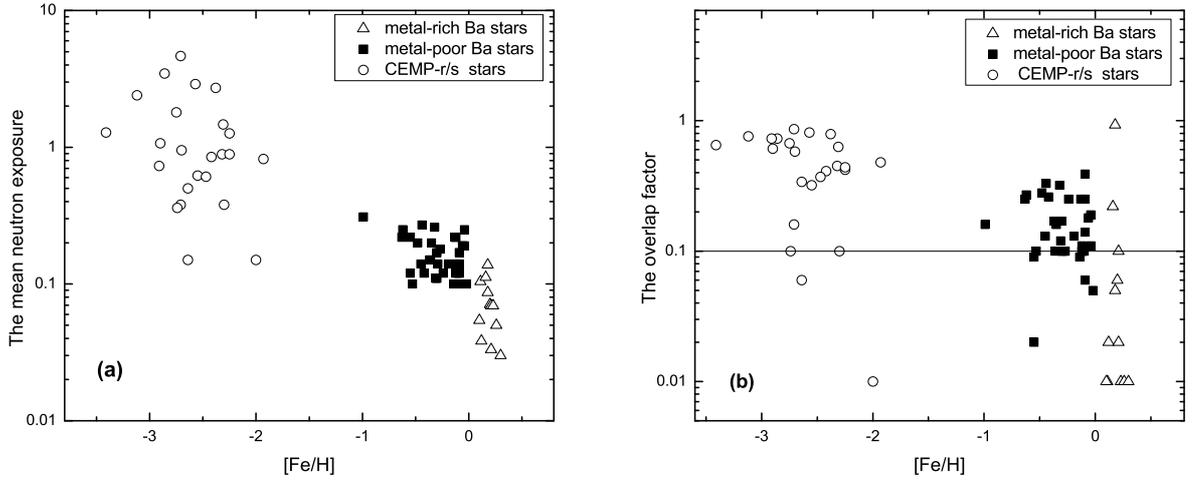}
  \vspace{0cm}
  \caption{Physical parameters  of s-process nucleosynthesis versus [Fe/H]: (a) for the mean neutron exposure,  $\tau_0$; (b) for the overlap factors, $r$. We adopted the same CEMP-r/s (open circles) and metal-poor barium (filled squares) star samples  as collected by \citet[][and see the references therein]{cui13}.  To use as homogeneous a sample as  possible,  we show only the sample presented in \citet{all06} for the metal-poor barium stars. The open triangles represent the metal-rich barium stars adopted from \citet{per11b}. \label{fig1}}
\end{figure*}

The mean neutron exposure, $\tau_0=-\Delta\tau/$ln$r$, is a fundamental parameter for the s-process nucleosynthesis, where $\Delta\tau$ is the neutron exposure per thermal pulse in  an AGB star, and $r$ is the fraction of material that remains to experience subsequent neutron exposures. Figure~\ref{fig1} shows $\tau_0$ and $r$ as a function of [Fe/H], where (a) for $\tau_0$, and (b) for $r$, respectively. We can see from Figure~\ref{fig1}a that $\tau_0$ decreases with increasing [Fe/H].  Because a higher $\tau_0$ value favors the production of heavier elements such as Ba, La,  and Pb \citep{bus01}, this means that the production of s-elements is less  efficient in metal-rich environments ([Fe/H]\,$>0$). In other words, the rule of the s-process nucleosynthesis summarized by \citet{gal98}  and \citet{bus01} is still valid  when it extends to metal-rich conditions. From Figure~\ref{fig1}b, we can see that more and more barium stars tend to have lower $r$ values with increasing metallicity. In our metal-rich barium stars,  almost all of the $r$ values are lower than 0.1, except for HD 130386 and HD 100012.  This supports the conclusion that  single neutron exposure events (usually with $r\le0.1$) for the s-process nucleosynthesis  probable exist  throughout the entire  metallicity range of the Galaxy evolution \citep{cui13}. Furthermore, the occurrence frequency of  single exposure events is very high in metal-rich conditions.

Table~\ref{tbl-1} shows that eight barium stars of  our sample of 12 metal-rich stars have  high $C_r$ values ($>5.0$)),  CD-25$\degr$ 6606, HD 46040, HD 49841, HD 84734, HD 85205, HD 100012, HD  101079, and HD 139660. This means that these barium stars have significant  r-process abundance characteristics, because the physical parameter, $C_r$, represents the r-process contribution to the neutron-capture elements (see formula 1). 
This is very confusing because considering their metallicities and  observational errors of the element abundances, the $C_r$ values should be  lower than about 4.0. In other words,  significant r-process abundance characteristics are not expected in metal-rich environments. Figure~\ref{fig2} shows the abundance ratios [La/Fe] versus [Nd/Fe]. In this figure, six metal-rich barium stars with  high $C_r$ are in the same region with the r-rich stars, and their [La/Fe] and [Nd/Fe] ratios also show a  highly positive correlation. It seems that these six metal-rich barium stars belong to an independent group  because they have similar abundance characteristics with the r-rich stars. For the metal-poor barium stars, however,  this is not the case \citep[][see their Table 1, 2]{cui13}. 

\begin{figure}[!ht]
  \centering
  \includegraphics[width=0.45\textwidth]{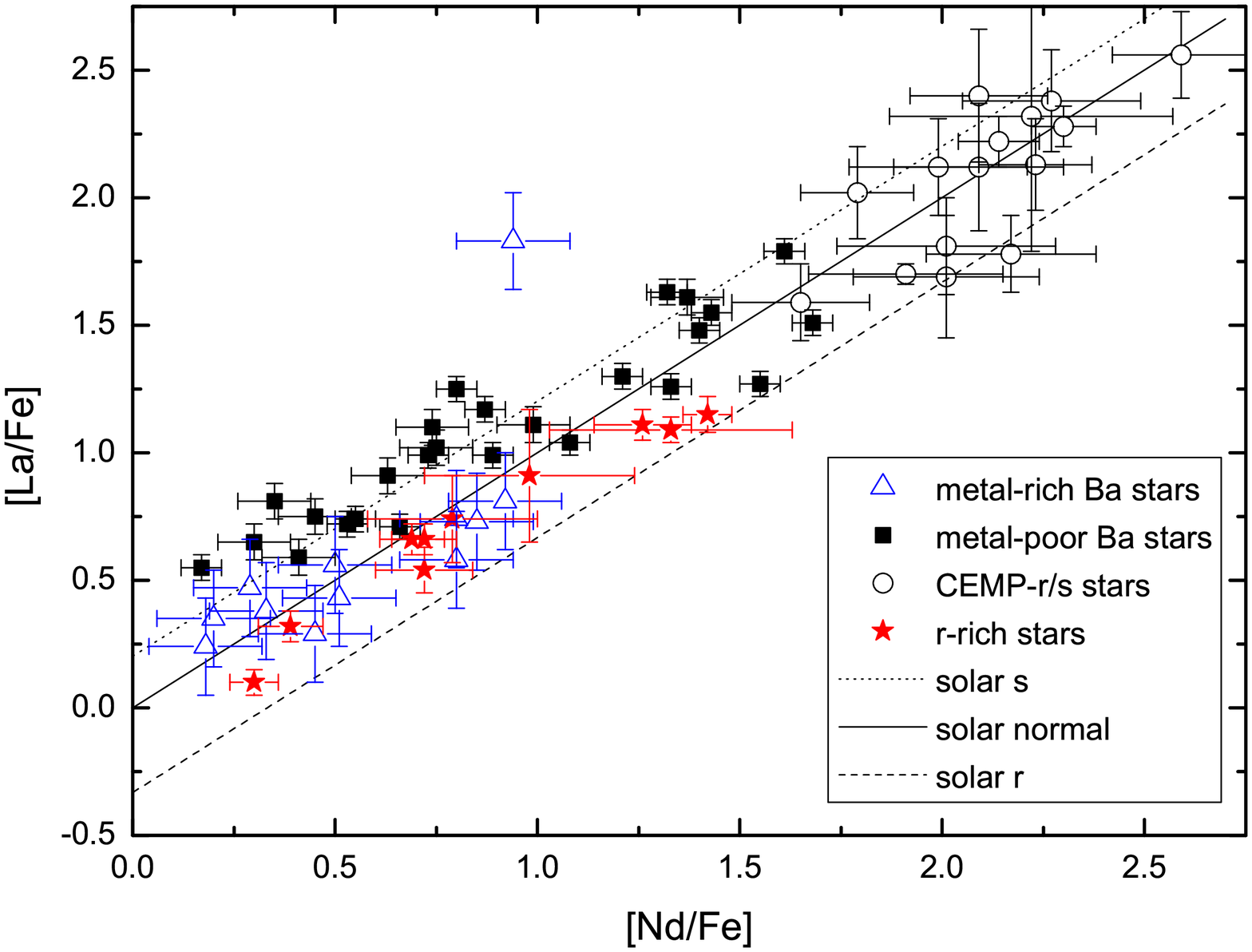}
 \vspace{0cm}
  \caption{  Abundance ratios [La/Fe] versus [Nd/Fe].  Error bars are also plotted. The symbols are the same as in Figure~\ref{fig1}. Here, r-rich stars (filled stars, including seven r-II stars and three r-I stars)  are also presented for comparison. The r-rich stars  are taken from the sample collected by \citet[][and see references therein]{cui13a}. The lines represent the solar [La/Nd] values: dotted for solar s, dashed for solar  r, and solid for solar normal.  See the online version for the color figure. \label{fig2}}
\end{figure}

The abundance ratios [Zr/Nd] vs. [La/Nd] are presented in  Figure~\ref{fig3} and  show that these six metal-rich barium stars with  high $C_r$ and  the r-rich stars are distributed in the same area, $-0.4<$\,[La/Nd]\,$<0$ and $-0.7<$\,[Zr/Nd]\,$<0$. This further supports that these barium stars belong to an independent group. It is interesting that the [Zr/Nd] ratio of the solar r-process is  higher than in some r-rich stars, such as CS 22892-052, which is a typical r-II star ([Eu/Fe]\,$>1$), where no  contribution from the weak r-process  is expected \citep{tra04, ish05, arc11}. The reason is that  solar Zr  is enriched by massive stars through the weak r- and s-process nucleosynthesis,  but the weak s-process does not work in extremely metal-poor environments \citep{the00}. Studying these stars  might shed some light on the neutron-capture process including both r- and s-process  nucleosynthesis, especially in metal-rich conditions.

\begin{figure}[!h]
  \centering
  \includegraphics[width=0.45\textwidth]{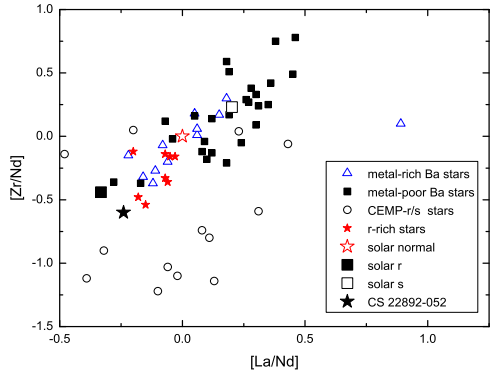}
  \caption{  Abundance ratios [Zr/Nd] versus [La/Nd]. The symbols are the same as in  Figures~\ref{fig1} and \ref{fig2}. The  large symbols represent solar values: filled square for solar r, open squares for solar  s, and open stars for solar normal, except  for the filled  star, which represents  the typical r-II star CS 22892-052.  See the online version for the color figure. \label{fig3}}
\end{figure}

Figure~\ref{fig4} shows the abundance ratios [La/Nd] as a function of [Fe/H] for metal-rich and metal-poor barium stars  and CEMP-r/s stars. Combing Figure~\ref{fig4} with Table~\ref{tbl-1}, we found that the metal-rich barium stars with  high $C_r$ values have a similar characteristic,  [La/Nd]\,$<0$, except for HD 139660. Because the [La/Nd] ratios can be used  as indicators of the relative contributions of s- and r-process nucleosynthesis for the heavy element productions in a star,  we call these barium stars with [La/Nd]\,$<0$ and $C_r>5.0$  r-rich barium  stars and the barium stars with [La/Nd]\,$>0$  normal barium stars.  Figure~\ref{fig4} shows that the  r-rich barium stars are spread in a similar [La/Nd] range  to those of the metal-poor r-rich stars and part of  the CEMP-r/s stars. Furthermore, even if the effects of different initial mass were considered, the AGB model yields of \citet{cri11} are also  unable to produce  negative [La/Nd] values  such as [Fe/H]\,$>-0.6$.  Thus, the  r-rich metal-rich barium stars seem to have same abundance behavior  as r-rich or CEMP-r/s stars. Of course,  more observations to measure the abundances of Ba and Eu are needed.

\begin{figure}[!ht]
  \centering
 \includegraphics[width=0.45\textwidth]{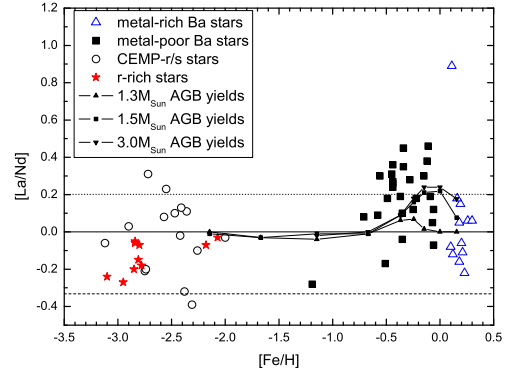}
  \caption{  Abundance ratios [La/Nd] versus [Fe/H]. The symbols are the same as in Figure~\ref{fig1}.  The small symbols represent the AGB model yields of \citet{cri11}.  See the online version for the color figure. \label{fig4}}
\end{figure}

Figure~\ref{fig4} also shows that the CEMP-r/s stars are spread in  a wide range of [La/Nd] about from $-0.4$ to 0.3.  Much effort has been  made to solve the problem of the origin of CEMP-r/s stars \citep{cui10,cui13,bis09}.  Commonly, a mixed origin  is assumed that the observed CEMP-r/s star had been  successively polluted by the s- and r-process  materials, although the accurate site for the r-process nucleosynthesis is still uncertain. In addition, the AGB models  also predict negative [La/Nd] values when different efficiencies of  the $^{13}$C neutron source are  included in the  calculations, especially in extremely metal-poor conditions ([Fe/H]\,$\le-2.0$) \citep{bus01, str06, zha06, bis10}, but  this does not work in metal-rich conditions \citep[][see their Figure 5, bottom panel]{bis10}. When a mixed contribution of the r- and s-process nucleosynthesis was considered, the observed abundance patterns of CEMP-r/s  stars were able to be fitted well by the theoretical predictions \citep{cui07, cui10, cui13, bis09}. This is also correct for almost all metal-poor and metal-rich barium stars including all  r-rich metal-rich barium stars. There are four metal-poor barium stars,  HD 749, HD 5424, HD  123396, and HD 223938, which have negative [La/Nd] values but  low $C_r$, they are only a small part of our 26 metal-poor barium star sample. None of them was able to be fitted well using the mixed mechanism. Two of them, HD 5424, HD 123396, have lower values of [La/Nd], $-0.17$ and $-0.28$, and also have  higher $\chi^2$, 10.52647 and 13.98755, respectively.  The reason  is unclear, however. 

R-rich stars are commonly suggested to be formed from  molecular clouds  that had been polluted by SNII. For the CEMP-r/s stars, the  favored mechanism is  pre-enrichment: the first formed r-rich star is polluted by the s-process material through mass-transfer from its former AGB companion in a binary system. But this was  challenged by \citet{lug12} because of the following  reasons: first, it is difficult to explain the linear correlation observed between [Ba/Fe] and [Eu/Fe] in CEMP-r/s  stars; the initial [r/Fe] does not affect the final [Ba/Fe] \citep{bis09}.  Second, it is difficult to understand that  there are fewer observed r-II stars  ([Eu/Fe]\,$>1$) than CEMP-r/s stars, if CEMP-r/s stars were suggested to evolve from r-II stars in binary systems. Finally, the metallicity distribution for the two groups is different \citep[see details in][]{lug12}. For these  reasons, \citet{lug12}  proposed a new r/s-process as a possible choice for CEMP-r/s star formation. This process  has features  that mix or superpose the s- and r-process.  This might be a choice for the origin of the  r-rich barium stars. However, the inhomogeneity of the interstellar medium is not expected in such a environment with [Fe/H]\,$>0$.  \citet{all12}   proposed CEMP-r/s stars have the same origin  as CEMP-s stars. In other words, the additional r-process site is not needed for the formation of CEMP-r/s stars. The reason  that the s-process  cannot explain the abundance patterns of the CEMP-r/s stars alone   is our incomplete knowledge of the Eu production by the s-process nucleosynthesis at low metallicity. It may be the most possible formation mechanism for the   r-rich barium stars. It is also very important to include Eu and other relevant elements in studying the metal-rich barium stars.

In this work, Nd and La abundances are very important to  determine whether a star belongs to the class of   r-rich barium stars. As a typical star of the sample,  \citet{per11b}  presented abundance uncertainties for the star CD-25$\degr$ 6606  (see their Table 7). Using the method presented by \citet{aok01}  and \citet{zha06}, we discuss the uncertainty of the model parameter  $C_r$, which is propagated from the abundance uncertainties of Nd and La.   Figure~\ref{fig5} shows the calculated abundance ratios [Nd/Fe] and [La/Fe] as a function of the r-process component coefficient $C_r$ in a model with $\Delta\tau=0.15$, $r=0.02$ and $C_s=0.00971$. These calculated results are compared with the observed abundances ratios of CD-25$\degr$ 6606. There is only one region of overlap in Figure~\ref{fig5}, $C_r=15.3^{+6.3}_{-2.0}$, in which the observed ratios of both [Nd/Fe] and [La/Fe] can be accounted  for well. The bottom panel displays the $\chi^2$ values calculated by the parametric model, and the minimum is $\chi^2=0.69281$ at $C_r=15.3$. It can be seen that the physical parameter $C_r$ can be constrained in a reasonable region.  We used $C_r\geq5.0$ as the reference value to judge whether a star  belonged to  the  r-rich barium star group. In fact, this value is  lower than 2.0 in  most normal barium stars \citep{cui13}. 

\begin{figure}[!ht]
  \centering
  \includegraphics[width=0.45\textwidth]{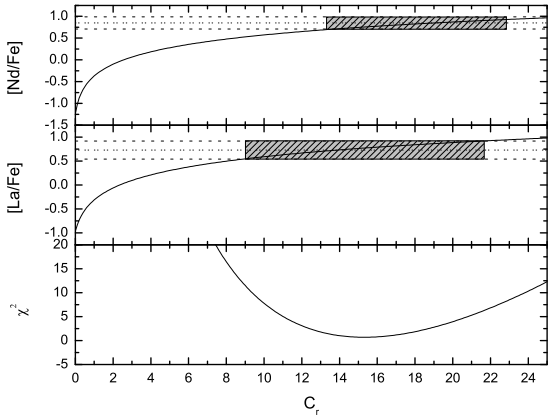}
  \caption{Abundance ratios [Nd/Fe] (top), [La/Fe] (middle), and $\chi^2$ (bottom), as a function of the r-process component coefficient in a model with $\Delta\tau=0.15$, $r=0.02$, $C_s=0.00971$. Solid curves refer to the theoretical results,  dotted horizontal lines  to the observational results with errors represented by dashed lines. The shaded area illustrates the allowed region for the theoretical model.  \label{fig5}}
\end{figure}

\section{Conclusions}\label{Sect:Conclusions}

Using a parametric model, we studied  the sample of metal-rich barium stars provided by \citet{per11b}. Our  results show that more than half of the sample stars need  high $C_r$ values to  be best fitted. In other words, these barium stars have a significant r-process characteristic. Based on this, we divided the barium stars into two groups: the  r-rich barium stars ($C_r>5.0$,  [La/Nd]\,$<0$)  and normal barium stars. The  r-rich barium stars seem to have the same abundance behavior  as the metal-poor r-rich and CEMP-r/s stars.  Although their abundance patterns  were fitted very well using  a mixed mechanism such as pre-enrichment, we still think that the most plausible formation mechanism for these stars is the s-process pollution.  That we  cannot explain them well using the s-process nucleosynthesis alone is maybe due to our incomplete knowledge on the production of Nd and other relevant elements by the s-process in metal-rich and super metal-rich environments \citep[see details in][]{all12}. 

Obviously, large uncertainties still remain in this topic, and a full understanding of these  r-rich barium stars will depend on the abundance information of Ba and Eu.  More in-depth theoretical studies of the s-process nucleosynthesis will  also help to shed light on the origin of  r-rich barium stars and CEMP-r/s stars.

\begin{acknowledgements}

 We acknowledge the  referee for helpful comments and suggestions  that significantly improved the paper. This work was  
 supported by  the National Natural Science Foundation of China under grant U1231119, 11273011, 11021504, 11390371, 11003002, 10973016, the China Postdoctoral Science Foundation under grant 2013M531587, the Natural Science Foundation of Hebei Province under grants A2011205102, A2011205067 and the Program for Excellent Innovative Talents in University of Hebei Province under grant CPRC034, the Natural Science Foundation of Hebei Provincial Education Department under grant Z2010168. 

\end{acknowledgements}

\end{document}